\documentclass[aps,prl,twocolumn,groupedaddress,showpacs]{revtex4-1}
\usepackage{amssymb}
\usepackage{latexsym}
\usepackage{graphicx}
\usepackage{enumerate}
\usepackage{amsmath}
\usepackage{dcolumn}
\usepackage{bm}
\usepackage{hyperref}  

\begin{document}

\author{Graciana Puentes$^{1,2}$, Ilja Gerhardt$^{1}$, Fabian Katzschmann$^{3}$, Christine Silberhorn$^{3}$, J\"{o}rg Wrachtrup$^{1}$, and Maciej Lewenstein$^{2}$}
\affiliation{$^{1}$$\mathrm{III}$ Physikalisches Institut, Research Center SCOPE, and MPI for Solid State Research, University of Stuttgart, Pfaffenwaldring 57, 70569 Stuttgart, Germany.\\
\noindent $^{2}$ ICFO - The Institute of Photonic Sciences, Mediterranean Technology Park,
Av. Carl Friedrich Gauss 3, 08860 Castelldefels, Barcelona, Spain.\\
\noindent $^{3}$ Applied Physics, University of Paderborn, Warburger Straße 100, 33098 Paderborn, Germany.}

\title{Observation of Topological Structures in Photonic Quantum Walks}

\begin{abstract}
Phases of matter with non-trivial topological order are predicted to exhibit a variety of exotic phenomena, such as  robust localized bound states in 1D systems, and edge states in 2D systems, which are expected to display  spin-helicity, immunity to back-scattering, and weak anti-localization. In this Letter, we present an experimental observation of topological structures generated via the controlled implementation of two consecutive non-commuting rotations in photonic discrete-time quantum walks. The second rotation introduces valley-like Dirac points in the system, allowing to create the non-trivial topological pattern. By choosing specific values for the rotations, it is possible to coherently drive the system between topological sectors characterized by different topological invariants. We probe the full topological landscape, demonstrating the emergence of localized bound states hosted at the topological boundaries, and the existence of extremely localized or delocalized non-Gaussian quantum states. Our results pave the way for the study of valley-polarization and applications of topological mechanisms in robust optical-device engineering. 

\end{abstract}


\maketitle


Phase transitions play a fundamental role in science, and in physics  
in particular. While classical phase transitions are typically driven  
by thermal noise, quantum phase transitions are triggered by quantum  
fluctuations \cite{Sachdev}.  Quantum phase transitions have received  
increasing attention within the realm of ultra-cold atom in optical  
lattices \cite{Lewenstein,Greiner}. Standard phase transitions, both  
classical and quantum, follow the, so called, Landau scenario, and  
consist in spontaneous symmetry breaking. The ordered phase can then  
be described by a \emph{local} order parameter.

A different kind of quantum phase transitions occurs in systems  
characterized by a, so called, topological order.  Such systems have  
generically degenerate ground states, which cannot be described by a  
local order parameter; they exhibit localized edge states, that are  
protected against noise by underlying symmetries. Paradigm examples of  
topological phases appear in quantum Hall effect (QHE)  
\cite{vKlitzing,Thouless}, fractional QHE \cite{Laughlin,Stormer}, and  
in spin QHE, or generally speaking in  topological insulators (TIs),  
predicted in  \cite{Kane,Bernevig} and realized experimentally in  
\cite{Koening,Hsieh,Qi,Chen,Xia}.

~Topological edge states characterizing TIs have recently been simulated in a number of different systems ranging from ultra-cold atoms in optical lattices  \cite{Esslinger}, to photonic networks of coupled resonators in silicon platforms \cite{Hafezi}.~Furthermore, it has recently become apparent that discrete-time quantum walks (DTQWs) \cite{Aharonov} offer a versatile platform for the exploration of a wide range of non-trivial topological effects (experiment) \cite{Kitagawa, Crespi, Alberti}, and (theory) \cite{Kitagawa2,Obuse,Shikano2, Asboth,Wojcik}. Further, QWs are roboust platforms for modelling a variety of dynamical processes from excitation transfer in spin chains \cite{Bose,Christandl} to energy transport in biological complexes \cite{Plenio}. They enable to study multi-path quantum inteference phenomena \cite{bosonsampling1,bosonsampling2,bosonsampling3,bosonsampling4}, and can provide for a route to validation of quantum complexity \cite{validation1,validation2}, and universal quantum computing \cite{Childs}. Moreover, multi-particle QWs warrant a powerfull tool for encoding information in an exponentially larger space, and for quantum simulations of biological, chemical and physical systems, in 1D and 2D geometries \cite{Peruzzo,Crespi,OBrien,Silberhorn2D}. 

\begin{figure} [t!]
\label{fig:1}
\includegraphics[width=1.0\linewidth]{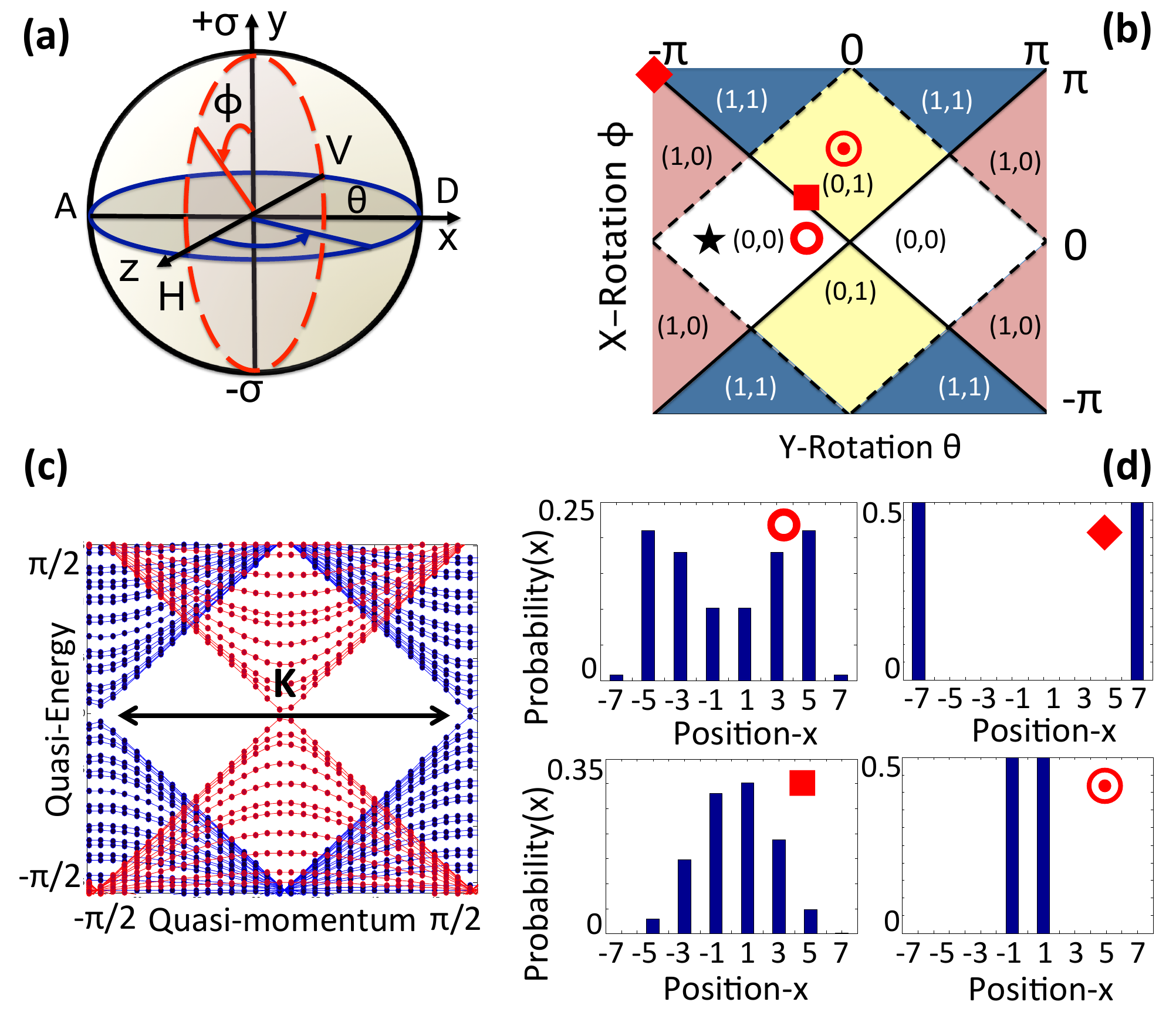} \caption{(a) Poincar\'{e} sphere representation, (b) phase diagram for parameter values within $-\pi \leq \theta, \phi \leq \pi$.  (c) Band structure in first Brillouin zone for rotation parameters $-\pi/2  \le \theta,\phi \le \pi/2 $. Red(blue) lines correspond $\phi=0$$(\pi/2)$. $\phi=\pm \pi/2$  allows to close the quasi-energy gap at additional zero quasi-energy points for quasi-momentum $k=\pm \pi/2$. Such Dirac points separated by the crystal momentum $K$ are analogous to valley-Dirac points in graphene. (d) Numerically simulated  benchmark states after $N=7$ steps. Circle: delocalized (non-Gaussian) quatum state ($\theta=\pi/4, \phi=0$), for Hadamard QW; Square: Localized bound state at topological boundary ($\theta=\pi/4, \phi=\pi/4$); Romboid: Extremely delocalized (non-Gaussian) quantum state at edge of Briulluin zone ($\theta=-\pi, \phi=\pi$); Filled circles: extremely localized state at center of topological sector ($\theta=0, \phi=\pi/2$).  }
\end{figure}

In this Letter, we present a novel theoretical and experimental scheme for generation of topological structure in 1D time-multiplexed DTQW architectures \cite{photons, Silberhorn}. The novelty relies on the introduction of two consecutive non-commuting rotations along the walk (Fig. 1 (a)). The second rotation allows to close the quasi-energy gap at additional points, and creates the non-trivial topological structure  (Fig. 1 (b)). In contrast to previous work \cite{Kitagawa}, the additional Dirac points are located at non-zero quasi-momentum, emulating valley-Dirac points in graphene, or related materials (Fig. 1 (c)). For graphene, symmetry allowed energy bands form two pairs of cones (or valleys) located at quasi-momenta $k=\pm \pi/2$. As a crystal momentum $K$ separates the two valleys, the valley degrees-of-freedom (DOF) are robust against slowly varying potentials and scattering and have promising applications in valleytronics, in addition to be considered an alternative subspace for encoding quantum information \cite{ValleyDirac1,ValleyDirac2,ValleyDirac3,Gunawan}. Our scheme is of interest since it allows to access the valley subspace in photonic systems, and study the interplay between valley and polarization DOF. By coupling these two subspaces it should be possible to access a higher-dimensional Hilbert space of the radiation field, and enable the study of more complex topological phases, such as the spin Hall phase \cite{Kane,Bernevig}.~Additionally, we characterize the topological landscape by the topological numbers $(Q_{0},Q_{\pi})$ \cite{Asboth}, thus extending upon previous theoretical approaches \cite{Kitagawa2}. In our scheme the second rotation modifies the dispersion relation and band structure (Fig. 1 (c)), a feature which modifies the spatial shape of the associated wave-functions. In contrast with the homogeneous case studied in Ref. \cite{Kitagawa}, this feature allows to observe localization effects without the need for a spatial inhomogeneity or artificial boundary, as numerically verified in Fig 1 (d). Furthermore, since the dispersion law is modified, the bound states reported here have not previously been observed. Moreover, for specific rotation parameters our scheme allows for the direct observation of extremelly (de)localized quantum states.

The basic step in the standard DTQW is given by a unitary evolution operator $U(\theta)=TR_{\vec{n}}(\theta)$, where $R_{\vec{n}}(\theta)$ is a rotation along an arbitrary direction $\vec{n}=(n_{x},n_{y},n_{z})$, given by $R_{\vec{n}}(\theta)=
\left( {\begin{array}{cc}
 \cos(\theta)-in_{z}\sin(\theta) & (in_{x}-n_{y})\sin(\theta)  \\
 (in_{x}+n_{y})\sin(\theta) & \cos(\theta) +in_{z}\sin(\theta)  \\
 \end{array} } \right) $, in the Pauli basis \cite{Pauli}. In this basis, the y-rotation is defined by a coin operator of the form  
$R_{y}(\theta)=
\left( {\begin{array}{cc}
 \cos(\theta) & -\sin(\theta)  \\
 \sin(\theta) & \cos(\theta)  \\
 \end{array} } \right) $ \cite{Pauli}. This is  
followed by a spin- or polarization-dependent translation $T$ given by 
$
T=\sum_{x}|x+1\rangle\langle x | \otimes|H\rangle \langle H| +|x-1\rangle \langle x| \otimes |V\rangle \langle V|,
$
 where $H=(1,0)^{T}$ and $V=(0,1)^{T}$ (Fig. 1 (a)).
The evolution operator for a discrete-time step is equivalent to that generated by a Hamiltonian $H(\theta)$, such that $U(\theta)=e^{-iH(\theta)}$ ($\hbar=1$), with $H(\theta)=\int_{-\pi}^{\pi} dk[E_{\theta}(k)\vec{n}(k).\vec{\sigma}] \otimes |k \rangle \langle k|$ and $\vec{\sigma}$ the Pauli matrices, which readily reveals the spin-orbit coupling mechanism in the system.~The quantum walk described by $U(\theta)$ has been realized experimentally in a number of systems \cite{photons,photons2,ions, coldatoms}, and has been shown to posses chiral symmetry, and display Dirac-like dispersion relation given by $\cos(E_{\theta}(k))=\cos(k)\cos(\theta)$. Here, we present localization effects given by the introduction of a second rotation along the x-direction by an angle $\phi$, such that  the unitarity step becomes $U(\theta,\phi)=TR_{x}(\phi)R_{y}(\theta)$, where $R_{x}(\phi)$ is given, in the same basis, by \cite{Pauli}:
\begin{equation}
R_{x}(\phi)=
\left( {\begin{array}{cc}
 \cos(\phi) & i\sin(\phi)  \\
i \sin(\phi) & \cos(\phi)  \\
 \end{array} } \right).
 \end{equation}
The modified dispersion relation becomes $\cos(E_{\theta,\phi}(k))=\cos(k)\cos(\theta)\cos(\phi)+\sin(k)\sin(\theta)\sin(\phi)$, where we recover the Dirac-like dispersion relation for $\phi=0$. We stress that since the dispersion relation characterizing our system differs from the homogenous case \cite{Kitagawa}, the bound states observed here have not previously been reported.
Dispersion relations within the first Brillouin zone for the homogenous system ($\phi=0$, red curves), and for the inhomogeneous system ($\phi=\pi/2$, blue curves) are plotted in Fig. 1 (c).~In both cases the system shows Dirac-like (linear) dispersion relation. The second rotation ($\phi \neq 0$) permits to close the gap at zero energy for complementary points, and allows to create non-trivial topological structure in the system.
\begin{figure} [t!]
\label{fig:1}
\includegraphics[width=1\linewidth]{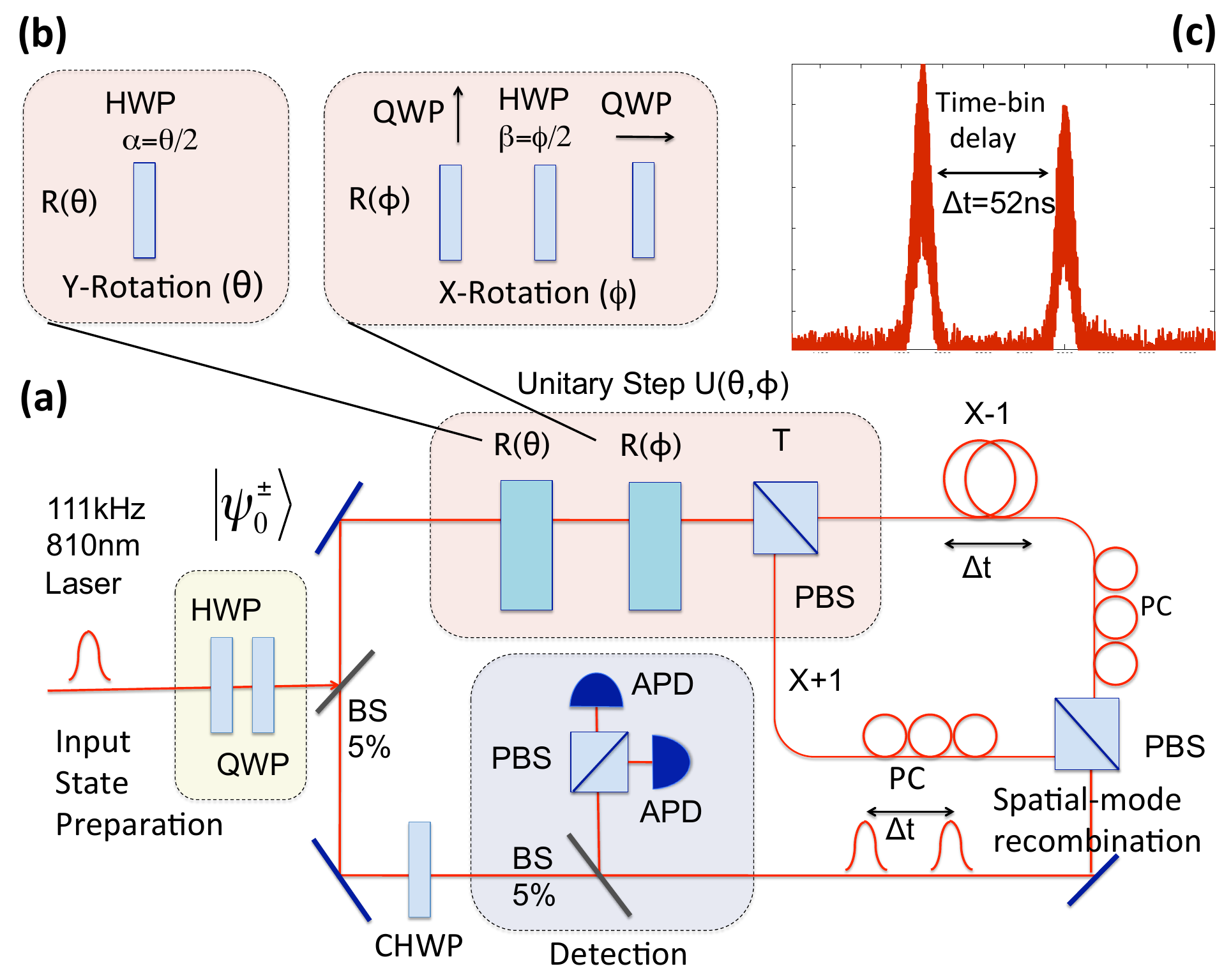} \caption{(a) Schematic of experimental setup. (b) Implementation of non-commuting rotations: $R_{y}(\theta)$ is implemented via a HWP at angle $\alpha=\theta/2$. $R_{x}(\phi)$ is implemented by a sequence of QWPs with fast axes oriented vertically and horizontally, respectively. In between the QWPs, a HPW oriented at $\beta=\phi/2$ determines the angle for the second rotation. (c) Histrogram of arrival times, after a trigger event at $t=0$.  }
\end{figure}
The topological phase diagram is depicted in Figure 1 (b).  Solid(dashed) lines join Dirac points where the gap closes for quasi-energy $E=0(\pi)$.~Valley-Dirac points correspond to $(\theta=\pm \pi/2, \phi=\pm \pi/2)$.~We characterize the modified topological landmark by determining the values for the topological invariants $(Q_{0},Q_{\pi})$  (where the subscript refers to the quasi-energy $E$), which correspond to the parity (even=0 or odd=1) of the number of times the gap closes at quasi-energy $E=(0,\pi)$ along a straight line in parameters space, starting from a fixed point in the zone $(0,0)$  \cite{Asboth}, indicated by a star. Note that such guideline does not refer to the actual trajectory followed by the quantum walker.~By choosing specific values for the rotation parameters $(\theta,\phi)$ it is possible to drive the system across topological sectors  characterized by different topological invariants $Q_{0,\pi}$. At the boundary of each topological sector the emergence of localized bound states is predicted. \\

\begin{figure} [b!]
\label{fig:1}
\hspace{-0.2cm}
\includegraphics[width=1.1\linewidth]{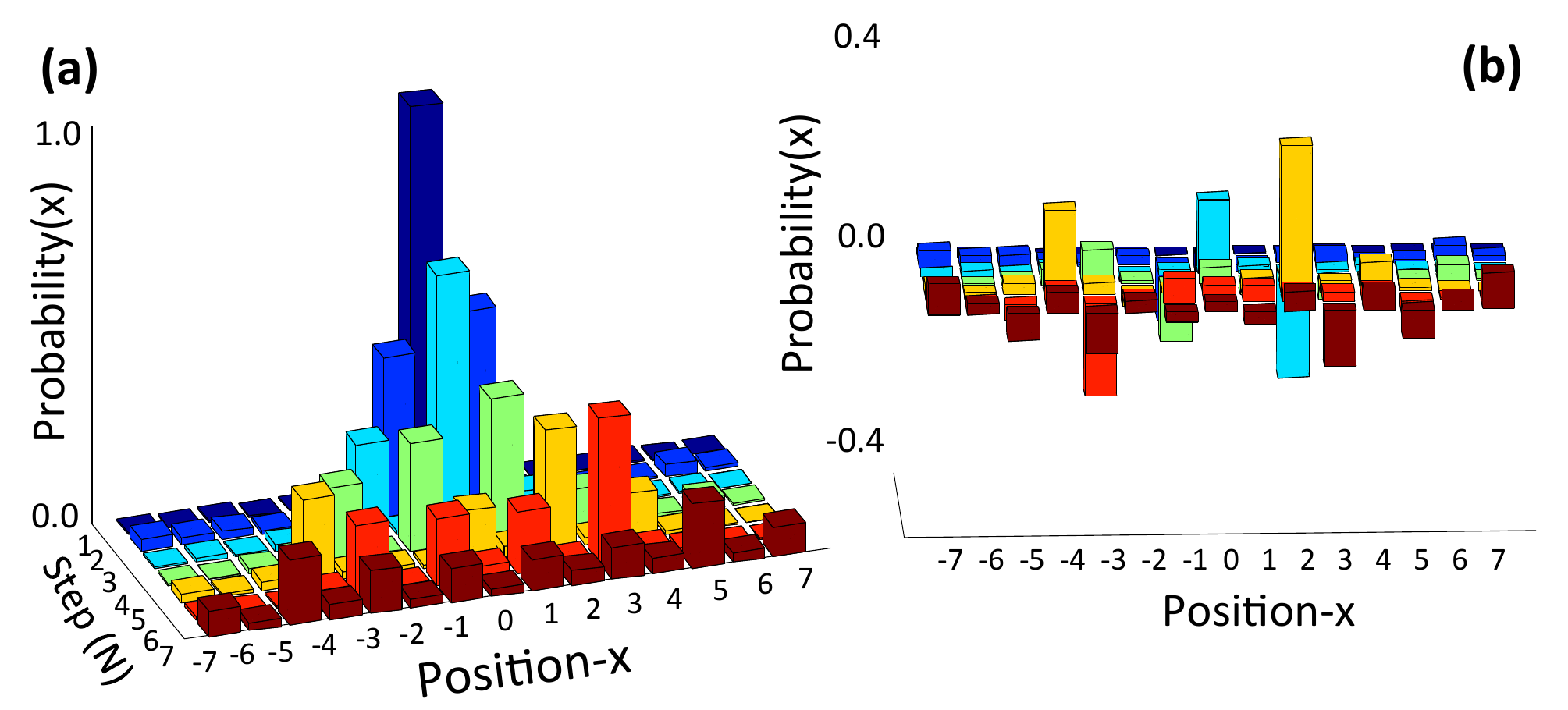} \caption{(a) Measured proability distributions for $N=7$ steps in  Hadamard QW  with $\theta=\pi/4, \phi=0$, and input state $|\psi_{0}^{+}\rangle$. (b) Difference between experiment and theory is within $20 \%$, and is mainly ascribed  to different soures of polarization dependent losses, spurious reflections, and shot-noise.}
\end{figure}

The experimental scheme for  time-multiplexed DTQWs with non-commuting coin operations is based on Ref.  \cite{Silberhorn} (see Fig. 2 (a)).  Our scheme allows to implement an arbitrarily large step-number in a compact architecture, in combination with detector gating and suitable ND filters.~Equivalent single-photon states are generated with an attenuated pulsed diode laser centered at 810 nm and with 111 kHz repetition rate (RR).  The initial state of the photons is controlled via half-wave plates (HWPs) and quarter-wave plates (QWPs),  to produce eigen-states of chirality $|\psi_{0}^{\pm}\rangle=|0\rangle \otimes 1/\sqrt{2}(|H\rangle \pm i|V\rangle)$. Inside the loop, the first rotation ($R_{y}(\theta)$) is implemented by a HWP with its optical axis oriented at an angle $\alpha=\theta/2$.~The rotation along the x-axis  ($R_{x}(\phi)$) is implemented by a combination of two QWPs with axes oriented horizontally(vertically), characterized by Jones matrices of the form 
$
\left( {\begin{array}{cc} 1 & 0  
\\ 0 & (-)i  \\ 
\end{array} } \right)$ (Fig. 2 (b)). In between the QWPs, a HWP oriented at $\beta=\phi/2$ determines the angle for the x-rotation. The spin-dependent translation is realized in the time domain via a polarizing beam splitter (PBS) and a fiber delay line, in which horizontally polarized light follows a longer path. The resulting temporal difference  between both polarization components corresponds to a step in the spatial domain  ($x \pm 1$). Polarization controllers (PC) are introduced to compensate for arbitrary polarization rotations in the fibers. After implementing the time-delay the time-bins are recombined in a single spatial mode by means of a second PBS and are re-routed into the fiber loops. After a full evolution the photon wave-packet is distributed over several discrete positions, or time-bins. The detection is realized by coupling the photons out of the loop by a beam sampler (BS) with a probability of 5$\%$ per step. Compensation HWPs (CHWPs) are introduced to correct for dichroism at the beam samplers (BS).~We employ two avalanche photodiodes (APDs) to measure the photon arrival time and polarization properties. The probability that a photon undergoes a full round-trip is given by the overal coupling efficiency  ($>70 \%$) and the overall losses in the setup resulting in $\eta= 0.50$.  The average photon number per pulse is controlled via neutral density filters and is below $\langle n \rangle <0.003$ for the relevant iteration steps  ($N=7$) to ensure negligible contribution from multi-photon events. 

We characterized the round-trip time (RTT=$750$ ns) and the time-bin distance (TBD=$52$ ns) with a fast Oscilloscope (Lecroy 640ZI, 4GHz). The RTT, and the laser RR determine the maximum number of steps that can be observerd in our system ($N_{\mathrm{max}}=12$).  Therefore $N_{\mathrm{max}}$ can be easily increased by adjusting these two design parameters.  Figure 2 (c), shows typical time-bin traces obtained from time-delay histogram recorded  with 72 ps resolution. The actual number of counts was obtained by integrating over a narrow window. We first implemented the Hadamard quantum walk, by setting $\theta=\pi/4$ and $\phi=0$. This is shown in Fig. 3 (a) for the first $N=7$ steps with no numerical corrections for systematic errors, after background subtraction. We compare the theoretical and experimental probability distributions via the similariy $S=[\sum_{x} \sqrt{P_{\mathrm{theo}}(x)P_{\mathrm{exp}}(x)}]^2$, with $S=0(1)$ for orthogonal(identical) distributions \cite{Silberhorn2D}, typically obtaining $S  \approx 0.85$. The difference between raw data and theory are displayed in Fig. 3 (b). Experimental errors can be explained in terms of asymmetric coupling, imperfect polarization-rotation compensation in the fibers, unequal efficiency in the detectors, and other sources of polarization dependent losses, in addition to shot-noise. Uncontrolled reflections are a main source of error. We removed this by subtracting the counts of the two APDs, and filtering peaks located at positions different from the RTT and the TBD during data analysis. 
\begin{figure} [t!]
\includegraphics[width=1\linewidth]{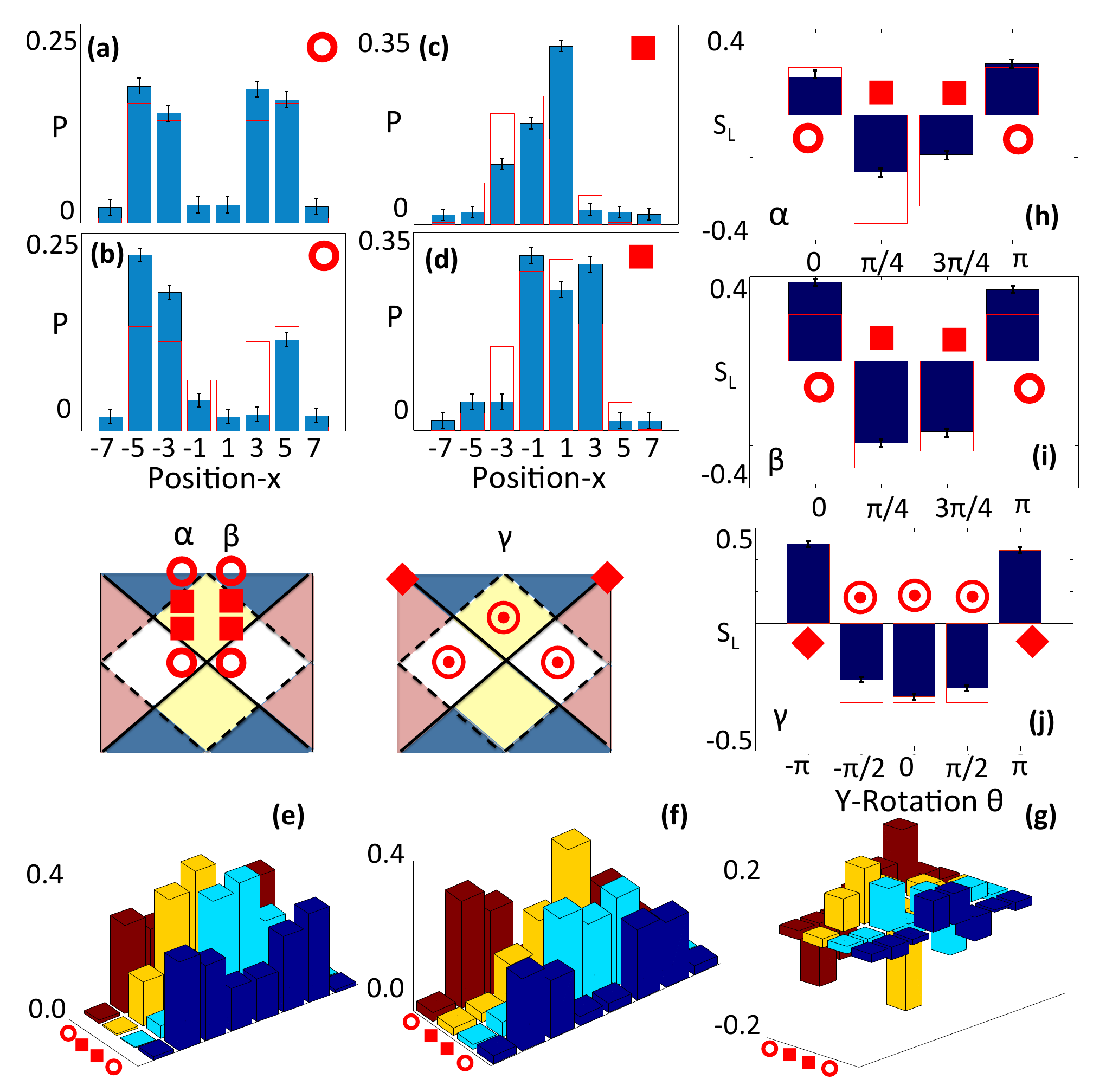} \caption{(a), (b), (c) and (d) Measured probability distributions (blue bars) after $N=7$ steps,  for a  trajectory characterized by $\theta=-\pi/4$ and $0 \le  \phi \le \pi $. Red empty bars correspond to numerical simulations. Figs. (a), (b) show delocalized  quantum states (circles) for $(\theta=-\pi/4,\phi=0,\pi)$. Figs. (c), (d) show localized bound states (squares) at the topological boundary $(\theta=-\pi/4,\phi=\pi/4,3\pi/4)$. Figs. (e), (f), and (g) numerical prediction,  experimental results and difference, respectively. (h) and (i) display measured  localization parameter (blue bars) $S_{L} = P_{\mathrm{outer}}-P_{\mathrm{inner}}$ for two symmetric trajectories ($\alpha$ and $\beta$). (j)  Experimental data points for extreme (de)localized states, saturating the bound of $S_{L}$ (trajectory $\gamma$}), for parameters ($\theta=\pm \pi,\phi=\pi$, romboids), and  ($\theta=(0, \pm \pi/2),\phi=(0,\pm \pi/2)$, filled circles). 
\end{figure}

Next, we probed the topological landscape (Fig. 4 inset) for two input states $|\psi_{0}^{\pm}\rangle$ of well defined chirality obtaining equlvalent results. Experimental results are only displayed for  $|\psi_{0}^{+}\rangle$. Figure 4 (a), (b), (c) and (d) show experimentally reconstructed probability distributions  (blue bars) for a  trajectory characterized by fixed $\theta=-\pi/4$ and $0 \le  \phi \le \pi $ after  $N=7$ steps, by tracing over polarization DOF. Red empty bars are numerical simulations, error bars are statistical. (a), (b) Delocalized quantum state for $(\phi =0, \pi)$, displaying characteristic non-Gaussian distribution corresponding to Hadamard QW; (c), (d), localized states for $(\phi =\pi/4, 3\pi/4)$ at topological boundaries. Fig. 4 (e), (f) and (g) display theoretical prediction, experimental results, and difference, repectively. In order to quantify the degree of localization we define a localization parameter ($S_{L}$) as the difference between outer and inner probability peaks $-1/2\le S_{L}=P_{\mathrm{outer}}-P_{\mathrm{inner}}\le 1/2$, located at  positions $|x_{\mathrm{outer}}|=5$ and $|x_{\mathrm{inner}}|=1$, for the $N=7$ step. We note that $S_{L}$ is not an order parameter \cite{Sachdev}, rather it is a simple way of characterizing the shape of the probability distributions. Localized(delocalized) quantum states are characterized by a negative(positive) parameter ($S_{L}<(>)0$). Experimentally reconstructed localization parameters ($S_{L}$) (blue bars) for  three full trajectories in parameter space ($\alpha$, $\beta$, and $\gamma$) are displayed in Fig. 4 (h), (i) and (j), respectively. Red empty bars correspond to theoretically expected values. Trajectories $\alpha$ and $\beta$ are symmetric as expected, and show localized states at the topological boundaries (squares), or delocalized non-Gaussian quantum states (circles) in the outer regions.  Additionally, we  demonstrate the existence of extremelly (de)localized states saturating the parameter bound $|S_{L}|=0.5$, as indicated in trajectory $\gamma$. While, strongly localized states are located at the center of each topological sector (full circles), strongly delocalized quantum states are found at the edges of the Brillouin zone (romboids). We measured the parameter ($S_{L}$) at four points displaying maximal (de)localization, confirming the complex topological structure in the system  (Fig.4). 

To conclude, we reported on a novel theoretical and experimental scheme for generation of topological structures in 1D  photonic DTQWs by tayloring two succesive non-commuting coin operations along the walk, experimentally confirming the existence of topological boundaries, localized bound states and extremelly delocalized non-Gaussian quantum states.~Localization effects are displayed after $N=7$ steps. This number could be  increased by adjusting the round-trip time and the laser repetition rate. Our scheme can be implemented for the study of  valley-polarization \cite{ValleyDirac1, ValleyDirac2, ValleyDirac3}. The results presented here can be generalized to 2D architectures \cite{Silberhorn2D}, and can find relevant applications in robust optical device engineering \cite{Hafezi2}, and entanglement topological protection \cite{MoulierasJPB}. \\

\emph{Acknowledgements.-} The authors gratefully acknowledge A. Schreiber, P. Neumann, F. Reinhard, Y. Shikano, J. Asb\'{o}th, J. P. Torres and S. Moulieras  for useful discussions and technical support. We acknowledge financial support by the Max-Planck-society, EU (Squtec), Darpa (Quasar), BMBF (CHIST-ERA), ERC (Quagatua, Osyris), and contract research of the Baden-W\"{u}rttemberg foundation.\\

\end{document}